\documentclass[aps,prl,twocolumn,showpacs,superscriptaddress,groupedaddress]{revtex4}
\usepackage{graphicx}
\usepackage{dcolumn}
\usepackage{bm}
\usepackage{amssymb}
\usepackage{amsmath}
\usepackage{amsthm}
\usepackage{epstopdf}
\usepackage{amsfonts}
\usepackage{verbatim}
\usepackage{feyn}
\usepackage{float}
\usepackage{sidecap}
\newcommand{\bi}{\boldsymbol{i}}
\newcommand{\bj}{\boldsymbol{j}}
\begin{document}

\widetext

\title{Particle Diagrams and Embedded Many-Body Random Matrix Theory}

\author{R.~A.~Small}\email{Rupert.Small@bristol.ac.uk}
\author{S.~M\"uller}
\affiliation{School of Mathematics, University of Bristol, Bristol BS8 1TW, UK}
\vskip 0.25cm
\date{\today}
\begin{abstract}
We present a new method which uses Feynman-like diagrams to calculate the statistical quantities of embedded many-body random matrix problems. The method provides a promising alternative to  existing techniques and offers many important simplifications. We use it here to find the fourth, sixth and eighth moments of the level density of a $m$-body system with $k$ fermions or bosons interacting through a random hermitian potential ($k\le m$) in the limit where the number of possible single-particle states is taken to infinity. All share the same transition, starting immediately after $2k=m$, from moments arising from a semi-circular level density to  gaussian moments. The results also reveal a striking feature; the domain of the $2n$'th moment is naturally divided into $n$ subdomains specified by the points $2k=m, 3k=m, \ldots, nk =m$.
\end{abstract}
\maketitle
\noindent\textit{Introduction.~}Since its  inception Random Matrix Theory (RMT) has found application in many domains of physics and even number theory\cite{akemann}\cite{guhr}. Subsequent attempts have been made to further refine the canonical form of RMT for many-particle systems\cite{french}\cite{bohigas}\cite{kota}. The need for such a refinement comes from the fact that canonical RMT explicitly assumes that each compound state interacts with every other, i.e., for a system of $m$ bodies the order of the interaction $k$ of the potential is equal to the number of particles $m$, in each quantum state. For many practical purposes however, we would like $k=2$, and ideally one should characterize the level density for the whole domain $1\le k\le m$. In this way one gains knowledge of how the statistics deform when going from the $k=2$ case to the canonical RMT ($k=m$) case.

The embedded ensembles, introduced by Mon and French\cite{mon} in 1975, gave physicists a powerful tool for extending RMT to the study of many-body interactions for the cases $k < m$. The method involves \textit{embedding} the \textit{k}-body potential ${V}_k$ into the \textit{m}-particle state space (see \cite{weidreview} and \cite{chaosreview} for reviews). At the turn of the last decade a breakthrough paper by Benet, Rupp and Weidenm\"uller \cite{weid} showed how a process of eigenvector expansions could be used to calculate certain statistical properties for embedded $k$-body potentials. These methods were difficult to implement however, and it remains unclear if they can practically be used to calculate moments higher than the fourth. By using a new method utilizing Feynman-like diagrams it becomes possible to do many calculations in a straightforward way, and we use the method here to find the fourth, sixth and eighth moments of the level density of the embedded Gaussian Unitary Ensemble (GUE). The method, which we will call the method of \textit{particle diagrams}, allows us to identify the order of magnitude of combinatorial expressions prior to calculating them explicitly. If we confine our interest to the usual limit case where the number of possible single-particle states $l \to \infty$ and $m \ll l$ this will in many cases provide a sufficient reason not to calculate certain terms at all, since we can foretell using particle diagrams that they will not survive in the asymptotic regime.

\begin{figure}[t]
\centering
\includegraphics[scale=.5]{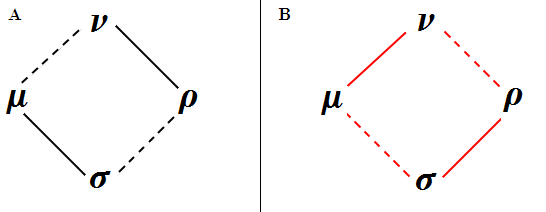}
\caption{\label{fig:q_term}(A) illustrates the particle diagram implied by  $A_{\mu\nu\rho\sigma}=\langle \mu|a_{\bj}^{\dag} a_{\bi}  |\sigma \rangle \langle \rho|a_{\bi}^{\dag}  a_{\bj} |\nu \rangle$ of (\ref{eq:ee46}) and (\ref{eq:ee52}), and (B) illustrates the particle diagram implied by the factor $A_{\sigma\mu\nu\rho}$ in (\ref{eq:ee52}). Each bond between compound states represents a set of single-particle states shared by both of the compound states.}
\end{figure}

\noindent\textit{The Embedded GUE.~}Consider the case of $m$ spinless fermions in a system with $l \gg m$ single particle levels, all interacting through a \textit{k}-body potential ($k \le m$) whose independent matrix elements follow a Gaussian probability distribution. Our single particle creation and annihilation operators are $a_j^{\dag}$ and $a_j$ respectively with $j = 1, \ldots, l$. We preserve notational traditions by writing the orthonormal $m$-particle states as $|\mu\rangle, |\nu\rangle, |\rho\rangle, etc.$ where each state takes the form $a_{{j_m}}^{\dag}\ldots a_{{j_1}}^{\dag}|0\rangle$ with $|0\rangle$ denoting the vacuum state and the restriction $1\leq j_1 < j_2 < \ldots < j_m\leq l$. The dimension of the state space is $N= {l\choose m}$ and the $k$-body potential is given by
\begin{equation}\label{eq:egue01}{V}_k = \sum_{\bj,\bi}v_{\bj\bi} a_{\bj}^{\dag}  a_{\bi}\end{equation}
where we abbreviate $\bj=(j_1,\ldots,j_k)$, $a_{\bj}=a_{j_k}\ldots a_{j_1}$ (similarly  for $\bi)$.
For the embedded GUE the only symmetry condition on the potential is that it be hermitian, namely $\langle \mu|V_k|\nu \rangle = \langle \nu|V_k|\mu \rangle ^{*}$ for all $\mu, \nu$. Matching coefficients gives ${v}_{\bj\bi}^{*} = {v}_{\bi\bj}$. As in canonical RMT ($k=m$) we suppose that matrix elements not related by hermitian symmetry are uncorrelated \textit{i.i.d} complex gaussian random variables with mean zero and variance $v_o^2$. Without loss of generality we take $ v_o^2 = 1$. For uncorrelated $v_{ \bj \bi}$ and $v_{\bj' \bi'}$ one has $\overline {v_{\bj\bi} v_{\bj'\bi'} } = 0$ whereas for $v_{\bj \bi} = v_{\bj{'}\bi{'}}^*$ the average instead becomes unity so that $\overline {v_{\bj \bi} v_{\bj{'} \bi{'}}} = \delta_{\bj \bi{'}}\delta_{\bi\bj^{'}}$. This equality is needed to calculate the average of powers of the trace of $V_k$. Of particular use in calculations is the abbreviation
\begin{align}\label{eq:ee46}A_{\mu\nu\rho\sigma} &:= \overline{\langle \mu|V_k|\sigma \rangle\langle \rho|V_k|\nu \rangle} =  \langle \mu|a_{\bj}^{\dag}  a_{\bi} |\sigma \rangle \langle \rho|a_{\bi}^{\dag}a_{\bj} |\nu \rangle\end{align}
closely related to the second moment of the level density. Here summation over the repeated indices $\bi, \bj$ is implied. For $A_{\mu\nu\rho\sigma}$ to be non-vanishing, $|\sigma\rangle$ and $|\rho\rangle$ must both contain the $k$ states included in $\bi$, and  $|\mu\rangle$ and $|\nu\rangle$ must both contain the $k$ states included in $\bj$. In addition $a_{\bi}|\mu\rangle$ and $a_{\bj}|\sigma\rangle$ have to contain the same states implying that $|\mu\rangle$ and $|\sigma\rangle$  coincide in the $m-k$ single-particle states not included in $\bi$ or $\bj$, and the same applies to $|\rho\rangle$ and $|\nu\rangle$. These relations are illustrated in Fig. \ref{fig:q_term}(A) where solid bonds $~\feyn{f}~$ connect many-particle states sharing $m-k$ single-particle states and dashed bonds $~\feyn{h}~$ connect many-particle states sharing $k$ single-particle states. Note that in the figure the overlaps indicated by neighboring bonds are disjoint, e.g., the states of $\bi$ form the overlap $\sigma\feyn{h}\rho $ but are excluded from the overlap $\mu\feyn{f}\sigma$. The ``particle diagrams'' drawn in this way form an essential ingredient for evaluating the moments of the level density.

With the odd moments being zero trivially and the second moment being used for normalization we start with the fourth moment, also called the kurtosis,\begin{equation}\label{eq:ee48}\kappa = \frac{\frac{1}{N}\mathrm{tr}(\overline{V_{k}^4})}{\left(\frac{1}{N}\mathrm{tr}(\overline{V_{k}^2})\right)^2}\;.\end{equation}
In the denominator and the numerator we have
\begin{eqnarray}\label{eq:ee49}
\mathrm{tr}(\overline{V_{k}^2}) &=& \sum_{\mu}\overline{\langle\mu|V_{k}^2|\mu\rangle} = A_{\mu\mu\rho\rho}\\
\label{eq:ee52}
\mathrm{tr}(\overline{V_{k}^4}) &=&  2 A_{\sigma\sigma\rho\rho}A_{\sigma\sigma\mu\mu} + A_{\mu\nu\rho\sigma}A_{\sigma\mu\nu\rho}
\end{eqnarray}
with the summations over repeated indices $\mu, \nu, \rho, \sigma$ implicit. One can see this using Wick contractions or from first principles by observing that the random variables ${v}_{\bj \bi}$ are gaussian so $\sqrt{\frac{\alpha}{\pi}}\int x^2 e^{-\alpha x^2}dx = \frac{1}{2\alpha}$ and $\sqrt{\frac{\alpha}{\pi}}\int x^4 e^{-\alpha x^2}dx = \frac{3}{4\alpha^2}$.
In (\ref{eq:ee49}) given the restrictions from Fig. \ref{fig:q_term}(A) we have to sum over all $|\mu\rangle$ and $|\rho\rangle$ sharing $m-k$ single-particle states.
There are $N = {l\choose m}$ states in the sum over all possible $|\mu\rangle$, ${m\choose m-k}$ ways to choose the overlap with $|\rho\rangle$, and ${l-(m-k)\choose k}$ ways to choose the rest of $|\rho\rangle$. Hence the result is
\begin{equation}\label{eq:arg3} \mathrm{tr}(\overline{V_{k}^2}) = {l \choose m}{m \choose k}{{l-m+k}\choose k}.\end{equation}
The trace in the numerator of $\kappa$ is given by (\ref{eq:ee52}). The calculation for the first term is almost identical, giving
\begin{equation}\label{eq:arg8} 2  A_{\sigma\sigma\rho\rho}A_{\sigma\sigma\mu\mu} = 2 {l \choose m}{m \choose k}^2{{l-m+k} \choose k}^2\end{equation}
(summation implicit) so that the only remaining term needed to complete the calculation for $\kappa$ is $A_{\mu\nu\rho\sigma}A_{\sigma\mu\nu\rho}$. This term,  as well as subsequent quotients defining the sixth and eighth moments, requires the summation of a series of binomial expressions not all of which are simple enough to write down, as we have done with (\ref{eq:arg3}) and (\ref{eq:arg8}). To decide which of these expressions survive in the limit of large  $l$ we define the {\it argument} of a binomial expression as its power in $l$ in the limit $l\to\infty$. For a quotient to give a nonvanishing result in that limit the argument of the numerator must be at least as large as the argument of the denominator. The argument can be obtained using Stirling's formula. Taking the dimension of the state space $N={l\choose m}$ as an example and applying Stirling's formula $lim_{n\to\infty}n! = \sqrt{2\pi n}{\left (\frac{n}{e}\right )}^n$ we take the value of the argument of $N$ as the power of $N(m!) \sim l^{l - (l - m)}$ which is $m$, so $\arg(N) = m$. More generally we have
\begin{equation}\label{eq:arg1}\arg\left [\prod_n{{l - a_n} \choose b_n}^{i_n}\right ] = \sum_n i_n b_n, \end{equation}
we also note that any additional factors independent of $l$ do not have any impact on the argument. Using  (\ref{eq:arg3}) it follows that $\arg [ ( \frac{1}{N} \mathrm{tr}(\overline{V_{k}^2}) )^2 ] = 2k$. Since we expect the value of the fourth moment to converge this gives us reason to predict that $\frac{1}{N}\mathrm{tr}(\overline{V_{k}^4})$ will have terms with arguments equal to $2k$  and possibly some terms with an argument less than $2k$. Those terms with an argument less than $2k$ will be ignored, contributing values of order no higher than $l^{-1}$ to $\kappa$ as $l\to\infty$, whereas those with argument equal to $2k$ must be calculated. Using (\ref{eq:arg8}) one easily sees that $2  A_{\sigma\sigma\rho\rho}A_{\sigma\sigma\mu\mu}$ has the argument $2k$ as expected (after division by $N$).

For the second summand $A_{\mu\nu\rho\sigma}A_{\sigma\mu\nu\rho}$ in (\ref{eq:ee52}) non-vanishing contributions arise if the indices of $A_{\mu\nu\rho\sigma}$ obey precisely the same restrictions as introduced earlier and depicted in Fig. \ref{fig:q_term}(A). The analogous restrictions to have non-vanishing $A_{\sigma\mu\nu\rho}$ are depicted in Fig. \ref{fig:q_term}(B). Figs. \ref{fig:q_term}(A) and (B) together form the particle diagram for $A_{\mu\nu\rho\sigma}A_{\sigma\mu\nu\rho}$, formally defined as the collection of bonds incorporating all restrictions for the indices. We are interested in the number of unique $m$-body states satisfying this diagram but we only need the term for which the argument reaches its maximal value $2k$. This is the term for which a maximal number of participating single-particle states can be chosen without restrictions; hence  the number of states participating in the bonds in Fig. \ref{fig:q_term} must be minimized. This means that e.g. the bond $\mu\feyn{h}\nu$ of Fig. \ref{fig:q_term}(A) must share the maximal number of states possible with the bond $\mu\feyn{f}\nu$ of Fig. \ref{fig:q_term}(B), implying that the larger of the two bonds contains all states included in the smaller one. This leaves $r:=\min(k,m-k)$ states participating in both bonds and $s:=|k-(m-k)|=m-2r$ states participating only in the bond involving more states. Analogous reasoning applies to each of the corresponding pairs of bonds of Fig. \ref{fig:q_term}(A) and Fig. \ref{fig:q_term}(B). The sets of $r$ overlapping states for all four such pairs may now coincide, and as we need to minimize the number of states participating in overlaps we are only interested in the case where they do coincide. On the other hand the set of $s$ states for each pair must be disjoint relative to the sets of $r$ overlapping states, as a consequence of neighboring bonds in the same diagram \ref{fig:q_term}(A) or \ref{fig:q_term}(B) being disjoint. Consequently the  possible choices for states are given by partitioning the $l$ available states into one set of $s$ states and four sets of $r$ states. To leading order we thus have the multinomial
\begin{align}\label{eq:arg11}&A_{\mu\nu\rho\sigma}A_{\sigma\mu\nu\rho} ={l\choose {s\; r\; r\; r\; r}}\nonumber\\
~~~~&= {l \choose m}{{l - m} \choose r}{{l - m - r}\choose r}{m \choose r}{{m - r}\choose r}\;.\end{align}
Recalling $N={l\choose m}$ the argument of the corresponding term in the numerator of $\kappa$ is then $\arg \left [ \frac{1}{N}\sum A_{\mu\nu\rho\sigma}A_{\sigma\mu\nu\rho}\right ] = 2r = 2\cdot min(k, m-k)$ so it is only for $k \le m-k$ that the argument of this term is equal to $2k$, while for $k > m - k$ it is always less. Finally, fitting all the surviving terms into the expression for $\kappa$ gives the limit form of the fourth moment as $l \to \infty$
\begin{align}\label{eq:arg16}\lim_{N\to\infty}\kappa = 2 + \lim_{N\to\infty}\frac{\frac{1}{N}A_{\mu\nu\rho\sigma}A_{\sigma\mu\nu\rho}}{\left [ \frac{1}{N}\sum A_{\mu\mu\rho\rho} \right ] ^2}
= 2 + \frac{{{m - k} \choose k}}{{m \choose k}}\end{align}
(with ${m-k\choose k}=0$ if $m-k<k$)
which corroborates the result found by Benet \textit{et. al.} \cite{weid} using the eigenvector expansion method and for $2k > m$ agrees with what is expected using the method of supersymmetry.

Although bosonic states can contain repeated single-particle states, any repeats will deplete the argument of the resulting binomial expressions so that the limit value of $\kappa$ for bosons is also given by (\ref{eq:arg16}). The same applies to subsequent expressions for the sixth and eighth moments. The reason behind this is the fact that the number of bosonic $m$-particle states containing repeats of $z\le m$ unique single-particle states is ${l\choose z}{{m-1}\choose{z-1}}$ which always has an argument less than $m$ except for when $z=m$; this leads to  bosonic $m$-particle states which contain no repeated single-particle states and are thus in one-to-one correspondence with fermionic states. A close correspondence between statistics for bosons and fermions for $l\to\infty$ is also noted in \cite{weidbose} where it is observed in particular for the fourth moment using different methods. As explained above however, this is a statement which can be extended to all moments. It should also be highlighted that particle diagrams analogous to those illustrated here can be drawn for traces arising in the embedded orthogonal and symplectic ensembles as well. Where Stirling's formula applies to the degrees of freedom implied by the bonds, these can likewise be used to calculate bounds on the order of magnitude.

\begin{figure}[t]
\centering
\includegraphics[scale=.5]{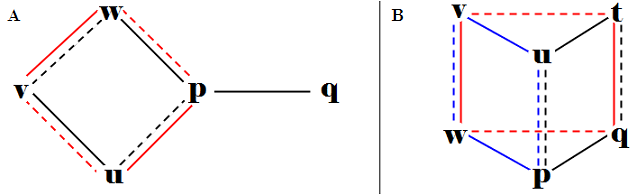}
\caption{(A) illustrates the particle diagram for the term $6A_{ptqq}\left (A_{twvu}A_{upwv} \right )$. This is almost identical to Fig. \ref{fig:q_term} but with the equivalent of Fig. \ref{fig:q_term}(A) and Fig. \ref{fig:q_term}(B) juxtaposed on the same diagram. The additional tail adds a factor ${m\choose k}{{l-m+k}\choose k}$. (B) illustrates the particle diagram of $3A_{putq}A_{qwvt}A_{upwv}$ where all the bonds implied by the expression are illustrated in a single 3D triangular prism. Using the diagram it becomes simpler to identify the single-particle states which must overlap maximally in order to maximise the argument of the whole sum.}
\label{fig:principle}
\end{figure}

\noindent\textit{The Sixth and Eighth Moments.~}Using the same method as above albeit with more complex particle diagrams, one can calculate the sixth and eighth moments of the level density as well. The sixth moment is given by
\begin{equation}\label{eq:s1}h = \frac{\frac{1}{N}\mathrm{tr}(\overline{V_{k}^6})}{\left(\frac{1}{N}\mathrm{tr}(\overline{V_{k}^2})\right)^3}.\end{equation}
Observing that $A_{\mu\nu\rho\sigma}^{} = A_{\sigma\mu\nu\rho}^{*}$ and $A_{\mu\nu\rho\sigma}^{} = A_{\rho\sigma\mu\nu}$,  and using Wick's theorem we obtain
\begin{align}\label{eq:s3} \frac{1}{N}\mathrm{tr}(\overline{V_{k}^6}) & = \frac{1}{N}\big[2~A_{ptqq}A_{tvuu}A_{vpww} + 3A_{ptqq}A_{uwvv}\left ( A_{tpwu} \right ) \notag\\
&+ 6A_{ptqq}\left (A_{twvu}A_{upwv} \right ) + 3A_{putq}A_{qwvt}A_{upwv} \notag\\
& + A_{pvuq}A_{qwvt}A_{tpwu}\big].\end{align}
Terms involving $A$'s with identical first and second (or third and fourth) indices simplify greatly as they give a contribution only if the two other indices coincide as well. For instance for $A_{ptqq}=\langle p|a_{\bj}^{\dag}a_{\bi}|q\rangle\langle q|a_{\bi}^{\dag} a_{\bj}|t\rangle$ to be nonzero the states $a_{\bj}|p\rangle$
and $a_{\bj}|t\rangle$ both have to coincide with $a_{\bi}|q\rangle$; adding the single-particle states with indices in $\bj$ then gives coinciding $|p\rangle$ and $|t\rangle$. Using this idea as well as the reasoning leading to (\ref{eq:arg3}) the first two terms in (\ref{eq:s3}) can be evaluated to give (after dividing out $N$) \begin{equation}\label{eq:s5}5\left[{m\choose k}{{l-m+k}\choose{k}} \right]^3.\end{equation}
For the third component $6A_{ptqq}\left (A_{twvu}A_{upwv} \right )$ we similarly require  $p=t$ for a non-zero contribution. The particle diagram for this term is illustrated in Fig. \ref{fig:principle}(A). Note that it is nearly identical to the particle diagram for (\ref{eq:arg11}) except for the addition of a tail ${p}\feyn{f}{q}$ which adds the factor ${m\choose k}{{l-m+k}\choose k}$ (obtained as for Eq. (\ref{eq:arg3})) to the expression we know already, so that the complete expression, after division by $N$, becomes
\begin{equation}{\scriptsize\label{eq:s6}6{m\choose k}{{l-m+k}\choose{k}}\left[{{l-m}\choose{r}}{{l-m-r}\choose r}{m\choose r}{{m-r}\choose r}\right]}.\end{equation}
Hence the argument of this term is $k+2r$ where $ r = min(k, m-k)$ is defined as before, so this term will only survive in the limit of $h$ as $l \to \infty$ for $k \le m-k$.

The particle diagram for the fourth term $A_{putq}A_{qwvt}A_{upwv}$ is illustrated in Fig. \ref{fig:principle}(B) as a single three dimensional triangular prism denoting the interrelated conditions on the states that must be satisfied for this term to be non-zero. The three faces correspond to the three factors. By the same principles as before adjacent bonds on the same face cannot share single-particle states. In order to maximise the argument for $2k \le m$ we choose the four sets of $k$ states between
 $|v\rangle$, $|t\rangle$, $|q\rangle$ and $|w\rangle$ as disjoint, afterwards choosing the $m-2k$ further states participating in $v\feyn{f}w$ but not in $v\feyn{h}w$. One can show that the choices made so far also uniquely determine the overlaps $v\feyn{f}w$ and $w\feyn{f}p$
on the `left' face in Fig.  \ref{fig:principle}(B). 
To fully determine $|u\rangle$ and $|p\rangle$ we then select the $k$ states participating in the `left' of the two bonds $u\feyn{h}p$,
altogether giving $m-k=m-2k+k$ states in addition to the original four sets of $k$ states. Considering the `right' face in an analogous way we obtain the same choice of $m-k$ states but broken down differently into a choice of $m-2k$ and a choice of $k$ states. Hence we have to consider all ways to select from $l$ given states four sets of $k$ states and one set of $m-k$ states, and then  split the latter into sets of $m-2k$ and $k$ states in two independent ways. This leads to
\begin{equation}\label{eq:s7}
{l \choose k\;k\;k\;k\;m-k}{m-k\choose k}^2 
\end{equation}
different choices.
For the final term $A_{pvuq}A_{qwvt}A_{tpwu}$ (not illustrated) one can show that its argument never exceeds $2m$ so that it will only contribute to the limit value of $h$ for $3k\le m$. Details of the calculation for this term and the eighth moment can be found in \cite{small}. The final result is
\begin{equation}\label{eq:s7.2}\frac{{\frac{1}{N}}A_{pvuq}A_{qwvt}A_{tpwu}}{\left(\frac{1}{N}\mathrm{tr}(\overline{V_{k}^2})\right)^3} = \frac{{{m-k}\choose{k}}{{m-2k}\choose{k}}}{{m\choose k}^2}.\end{equation}
Taking the quotient for $h$ using the above expressions gives the final result for the sixth moment
\begin{equation}\label{eq:s8} {\lim_{N\to\infty}}h = 5 + \frac{{{m-k}\choose{k}}{{m-2k}\choose{k}}}{{m\choose k}^2} + 6\frac{{{m-k}\choose{k}}}{{m\choose k}} + 3\frac{{{m-k}\choose{k}}^2}{{m\choose k}^2}.\end{equation}
For the eighth moment we again have to calculate products of ensemble averaged pairs of the matrix elements of ${V_k}$ however there are now $\frac{n!}{2^{n/2} (\frac{n}{2})!} = 105~~(n=8)$ components in the sum. As before, many of the particle diagrams can be expressed in the same way as diagrams we have already seen but with tails leading to additional combinatorial factors as in (\ref{eq:s6}). When selecting for the largest argument some particle diagrams also collapse to more familiar diagrams that have already been calculated for the lower moments. This becomes of great use in calculations, the details of which we present in \cite{small}. The final form for the eighth moment is
\begin{align}\label{eq:s11} &{\lim_{N\to\infty}}\tau = 14 + \frac{{{m-k}\choose{k}}{{m-2k}\choose{k}}{{m-3k}\choose{k}}}{{m\choose k}^3} + 4\frac{{{m-k}\choose{k}}{{m-2k}\choose{k}}^2}{{m\choose k}^3} \notag\\
&~~~~~~~+ 8\frac{{{m-k}\choose{k}}{{m-2k}\choose{k}}}{{m\choose k}^2} + 8\frac{{{m-k}\choose{k}}^2{{m-2k}\choose{k}}}{{m\choose k}^3} + 12\frac{{{m-k}\choose{k}}^3}{{m\choose k}^3} \notag\\
&~~~+ 28\frac{{{m-k}\choose{k}}^2}{{m\choose k}^2} + 28\frac{{{m-k}\choose{k}}}{{m\choose k}} + 2 \frac{{{m-k}\choose{k}}^2}{{m\choose k}^3} \sum_{\alpha} \frac{{{k}\choose{\alpha}}^2{{m-2k}\choose{k-\alpha}}}{{{m-k}\choose{\alpha}}}.\end{align}
\begin{figure}[h]
\centering
\hspace*{.05in}
\includegraphics[scale=.5]{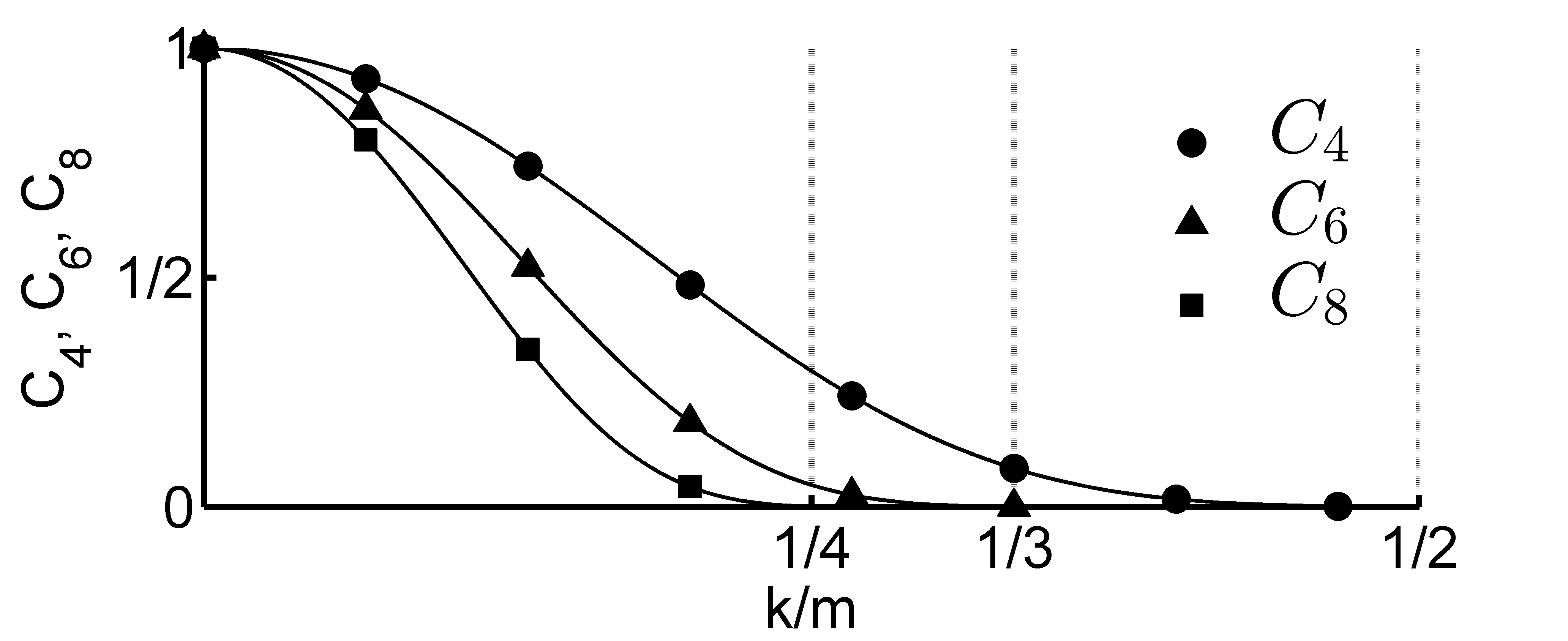}
\caption{Plot of the factors $C_4 = {{{m-k}\choose k}}/{{m\choose k}}$, $C_6 = {{{m-2k}\choose k}}/{{m\choose k}}$ and $C_8 = {{{m-3k}\choose k}}/{{m\choose k}}$ for $m=15$. The term with coefficient 1 for the fourth moment is $C_4$, for the sixth moment is $C_4 C_6$ and for the eighth moment is $C_4 C_6 C_8$. These coefficients illustrate the increasing degree to which the domain of the $2n$'th moment is partitioned for increasing $n$. For $k/m > \frac{1}{2}$ the moments are those of a semi-circle and for $k=0$ they are those of a gaussian.}
\label{fig:divisions}
\end{figure}

\noindent\textit{Conclusions.~}We have introduced a measure based on Stirling's formula as well as Feynman-like \textit{particle diagrams} to calculate statistics of embedded $k$-body random matrix potentials. We have illustrated the general method and supported its soundness by calculating and confirming the known expression for the fourth moment of the level density for the embedded GUE. Furthermore, we have shown the strength of these tools by calculating the sixth and eighth moments of the level density as well. The results reveal that certain behaviors identified with the fourth moment follow to at least the eighth moment and plausibly to all higher moments; a transition from a semi-circular moment ($\kappa = 2, h=5, \tau = 14$) starting immediately after $k=m/2$ to a gaussian moment ($\kappa = 3, h=15, \tau = 105$) for $k=0$. The results are also consistent with \cite{mon}, giving the expected gaussian moments in the dilute limit $k \ll m\ll l$. We have also shown that the domain of the $2n$'th moment manifests an interesting feature, namely a natural division at the points $2k = m, 3k=m, \ldots, nk=m$. We note the appearance in the eighth moment of a Hahn polynomial. The precise role of Hahn polynomials in the theory of Many-Body RMT is an open question. We have noted that by using $arguments$ and the method of particle diagrams it becomes self-evident that the limit values of the moments for the embedded GUE for fermions and bosons will always be equal. We believe that these methods hold promise for the further application of RMT to the study of many-body statistics; in spite of recent progress such as \cite{weidmetal}, a fully analytical study of the statistics in the embedded random matrix ensembles still poses a considerable challenge\cite{weidreview}\cite{srednicki}.

\end{document}